\documentclass[fleqn,usenatbib]{mnras}

\usepackage{newtxtext,newtxmath}

\usepackage[T1]{fontenc}
\usepackage{ae,aecompl}

\usepackage{graphicx}	
\usepackage{amsmath}	
\usepackage{amssymb}	

\title[Seeing above the Clouds]{Seeing above the Clouds with High Resolution Spectroscopy}

\author[Gandhi, Brogi and Webb]{
Siddharth Gandhi,$^{1,2}$\thanks{E-mail: Siddharth.Gandhi@warwick.ac.uk}
Matteo Brogi,$^{1,3,2}$
Rebecca K. Webb$^{1,2}$
\\
$^{1}$Department of Physics, University of Warwick, Coventry CV4 7AL, UK\\
$^{2}$Centre for Exoplanets and Habitability, University of Warwick, Gibbet Hill Road, Coventry CV4 7AL, UK\\
$^{3}$INAF - Osservatorio Astrofisico di Torino, Via Osservatorio 20, I-10025 Pino Torinese, Italy
}

\date{Accepted XXX. Received YYY; in original form ZZZ}

\pubyear{2020}

\begin{document}
\label{firstpage}
\pagerange{\pageref{firstpage}--\pageref{lastpage}}
\maketitle

\begin{abstract}
In the last decade ground based high resolution Doppler spectroscopy (HRS) has detected numerous species in transiting and non-transiting hot Jupiters, and is ideally placed for atmospheric characterisation of warm Neptunes and super Earths. Many of these cooler and smaller exoplanets have shown cloudy atmospheres from low resolution near infrared observations, making constraints on chemical species difficult. We investigate how HRS can improve on these given its sensitivity to spectral line cores which probe higher altitudes above the clouds. We model transmission spectra for the warm Neptune GJ~3470~b and determine the detectability of H$_2$O with the CARMENES, GIANO and SPIRou spectrographs. We also model a grid of spectra for another warm Neptune, GJ~436~b, over a range of cloud-top pressure and H$_2$O abundance. We show H$_2$O is detectable for both planets with modest observational time and that the high H$_2$O abundance-high cloud deck degeneracy is broken with HRS. However, meaningful constraints on abundance and cloud-top pressure are only possible in the high metallicity scenario. We also show that detections of CH$_4$ and NH$_3$ are possible from cloudy models of GJ~436~b. Lastly, we show how the presence of the Earth's transmission spectrum hinders the detection of H$_2$O for the most cloudy scenarios given that telluric absorption overlaps with the strongest H$_2$O features. The constraints possible with HRS on the molecular species can be used for compositional analysis and to study the chemical diversity of such planets in the future.
\end{abstract}




\section{Introduction}

Since the dawn of exoplanet science it was proposed that clouds could play an important role in their atmospheres \citep{Burrows1999}. Even at the high temperatures of hot Jupiters ($> 1000$~K), a number of species cross the condensation boundary, most notably silicate species such as enstatite (MgSiO$_3$), fosterite (Mg$_2$SiO$_4$), and spinel (MgAl$_2$O$_4$). The discovery of transiting planets \citep{Charbonneau2000, Henry2000} opened up the characterisation of exoplanet atmospheres, and clouds were immediately identified as having a dramatic effect on the amplitude and visibility of spectral features. \citet{Seager2000} proposed that observations of transmission spectra could be used to discriminate between atmospheric models, and \citet{Brown2001} provided analytic calculations showing how the transmission spectrum would look like in the presence of clouds. \citet{Sudarsky2000} proposed that the altitude at which condensate formed could impact the reflectivity of hot Jupiters in the optical, resulting in very high albedo for high silicate layers, and poorly reflective atmospheres for silicates forming below the photosphere, due to the presence of broad-band absorption from the doublets of alkali lines (K and Na).

Despite this theoretical support, the observational evidence for cloudy atmospheres remained essentially confined for many years to one exoplanet, HD~189733~b, for which early observations with the ACS camera on-board the Hubble Space Telescope \citep[HST,][]{Lecavelier2008, Pont2008} showed a sloped transmission spectrum that was interpreted as due to Rayleigh scattering from some source of broad-band opacity. Rather than clouds produced by condensation, the authors of these studies advocated for hazes, produced through photo-chemical reaction and proposed as possible source of opacity a few years earlier by \citet{Fortney2005}. The sloped optical transmission spectrum of HD~189733~b was subsequently confirmed and refined with HST STIS (Space Telescope Imaging Spectrograph) observations \citep{Sing2011}, and tentatively observed to extend to the near-infrared wavelengths covered by HST WFC3 (Wide Field Camera 3) \citep{Gibson2012}. The latter claim was subsequently revised by \citet{mccullough2014}, who detected water vapour absorption with the same instrument. They proposed that the sloped spectrum observed could instead be the product of uncorrected stellar activity, more specifically the spectrum of unocculted star spots that would change the effective stellar spectrum during and out of transit. Supporting the hazy scenario, there was also a tentative detection of polarised light from this planet \citep{Berdyugina2011} and the prediction of a high planet albedo, subsequently confirmed by the detection of a secondary eclipse of the planet in the blue optical \citep{Evans2013}.

The scarce evidence for clouds or hazes in exoplanet atmospheres changed with the observations of cooler Neptune or sub-Neptune transiting exoplanets, with the most compelling example being GJ~1214~b. Since its discovery by \citet{Charbonneau2009}, it was identified as one of the most promising targets for transmission spectroscopy, due to the low density and smaller host star (an M5.5 dwarf). However, progressively deeper and more precise observations showed a featureless spectrum \citep{bean2010,berta2012,kreidberg2014_gj1214}. The precision of the data is so stringent, that the only scenario compatible with the observed spectrum is a layer of thick clouds at high altitude suppressing all the atmospheric spectral features.

Since these pioneering observations the evidence for clouds or hazes has grown and become more widely recognised. The development of retrieval algorithms has allowed robust statistical significances to be placed on detections of hot Jupiters under transmission \citep[e.g.][]{madhu2009, madhu2014, evans2016, pinhas2018, wakeford2018, chachan2019, chubb2020}. This has produced an overwhelming consensus that clouds are indeed influencing measured exoplanet spectra and reducing the features of species such as H$_2$O \citep[e.g.][]{sing2016, barstow2017, pinhas2019}. For such hot planets it has also been possible to detect H$_2$O in retrievals of the secondary eclipse \citep[e.g.][]{crouzet2014, kreidberg2014, line2016, mikal-evans2020, lothringer2020} in addition to constraints on some other species \citep[e.g.][]{haynes2015, sheppard2017, arcangeli2018, mikal-evans2019, gandhi2020b}. 

Further evidence for clouds is also confirmed in the expanding sample of observations of super Earths and warm Neptunes \citep[e.g.][]{knutson2014, knutson2014_hd97658, wakeford2017, benneke2019, kreidberg2020}, few of which have revealed the presence of H$_2$O but many show muted or featureless spectra due to thick high altitude clouds at the terminator, making characterisation of the atmosphere difficult. Whilst some secondary eclipse constraints have been possible \citep{stevenson2010}, the emission spectra of cooler planets with equilibrium temperatures $\lesssim$1000~K are difficult to obtain at a sufficient level of precision. With surveys such as TESS \citep{ricker2015} finding more transiting planets we are likely to see more cloudy atmospheres in follow up observations of such cool planets. Thus characterising their atmosphere remains a significant challenge.

In this paper we explore the effects and detectability of clouds with high resolution spectroscopy (HRS), i.e. at spectral resolving powers $R\gtrsim25,000$. Exclusive dominion of ground-based observatories, HRS is rapidly evolving since its first successful inception \citep{snellen2010} and has been used to characterise the atmospheres of a growing number of transiting as well as non-transiting exoplanets \citep[see e.g. review by][]{birkby2018}. In the near infrared, detections in dayside and transmission spectra have been made for species such as H$_2$O \citep[e.g.][]{birkby2013,lockwood2014,Piskorz2016, Piskorz2017, birkby2017, webb2020} and CO \citep[e.g.][]{snellen2010, brogi2012, dekok2013, brogi2013, brogi2017}. In addition, HCN and CH$_4$ have been detected using instruments such as CRIRES and GIANO \citep{hawker2018, cabot2019, guilluy2019}. In the optical, refractory species such as TiO, VO and FeH have been observed \citep{nugroho2017, bourrier2020} in addition to atomic species such as Fe, Ti and Na \citep[e.g.][]{wyttenbach2015, louden2015, hoeijmakers2019, seidel2019, Gibson2020, cabot2020, ehrenreich2020}. These detections are of great importance given that such species with strong optical opacity are predicted to cause thermal inversions on the dayside \citep[e.g.][]{hubeny2003, fortney2008, molliere2015, lothringer2018, gandhi2019b}. In recent years the development of high resolution instrumentation with wide spectral range and high efficiency has extended HRS to smaller telescope facilities \citep[e.g.][]{follert2014, artigau2014, quirrenbach2014, park2014, rayner2016}. H$_2$O has already been observed on transiting hot Jupiters through some of these instruments \citep{brogi2018, alonso-floriano2019, sanchez-lopez2019}, and in the future this will also allow us to characterise smaller and cooler planets with weaker spectral signatures.

Detections of species with ground based HRS operates conceptually differently from low resolution observations of exoplanetary atmospheres, such as those with the HST WFC3 spectrograph mentioned above. Low resolution observations constrain species through their broad spectral bands of opacity over the continuum and are thus prone to degeneracies between clouds and species with overlapping bands \citep[e.g.][]{welbanks2019}. Detections with HRS on the other hand are achieved through the correlation of numerous ($\sim$10$^3$) transition lines in the spectrum from the planet. These planetary lines are Doppler shifted as the planet moves along the orbit, allowing us to extract the weaker planet signal over the stellar noise and the spectrum of the Earth's atmosphere. Due to the ability to resolve individual lines, HRS can robustly discriminate between species. This does however lose sensitivity to the broad absorption bands of each species as well as absolute eclipse depth (for emission spectroscopy) due to the particular algorithms used to remove unwanted spectral components. However, \citet{brogi2019} and \citet{Gibson2020} have shown that when HRS observations are placed into a Bayesian framework, the sensitivity to absolute abundances is still present in the data. This is because the log-likelihood metric used in these works preserves the relative depth and shape of spectral lines, which encodes the chemical and physical conditions of the atmosphere.

Importantly for this work, HRS also enhances the dynamic range in atmospheric pressures probed by spectroscopic observations. The opacity of the atmosphere changes by orders of magnitude between wavelengths in the core of a strong spectral line and wavelengths away from it. This in turns means that for high opacity regions the high resolution spectrum is formed much higher up in the atmosphere than the low-resolution spectrum. Thus HRS has the potential to probe above the clouds and thereby constrain the atmospheric abundances of such cloudy exoplanets. In addition, HRS has the potential to break the high metallicity/high altitude cloud deck degeneracy seen with many low resolution observations of cloudy exoplanets \citep[e.g.][]{knutson2014, kreidberg2014_gj1214}.

We demonstrate the feasibility of HRS to characterise the atmospheres of cloudy exoplanets by modelling two warm Neptunes observed to have cloudy atmospheres from low resolution observations, GJ~436~b and GJ~3470~b \citep{knutson2014, benneke2019}. We model varying composition and clouds and explore the detectability and abundance constraints of various volatile species, most notably H$_2$O. We first explore how the latest generation of facilities may detect H$_2$O on GJ~3470~b, which shows evidence of the species as well as a cloud deck from HST and Spitzer observations \citep{benneke2019}. We then model the warm Neptune GJ~436~b, which has shown a featureless transmission spectrum in the HST WFC3 range \citep{knutson2014}, and explore a grid over H$_2$O abundance and cloud deck to explore the constraints on H$_2$O as cloud pressure varies. In addition, we show that the degeneracy between high metallicity and high altitude clouds, which both result in low resolution spectra with muted features, may be broken with HRS. Finally, we explore how HRS may be used to detect other volatiles such as CH$_4$ and NH$_3$ and how these detection significances vary with cloud.

The next section discusses the transmission models and the simulated datasets for each instrument. We then describe the results and discussion in section \ref{sec:results}. This is followed by the conclusions in section \ref{sec:conclusion}.

\section{Methods}\label{sec:methods}

\subsection{Model Spectra}

We generate the spectral models using the GENESIS modelling framework \citep{gandhi2017} adapted for transmission spectroscopy \citep[e.g.][]{pinhas2018}. These models encompass a wide range of atmospheric composition for the various molecular species which may be present. We model the spectra on a grid of pressures spanning 10$^1$-10$^{-8}$~bar, assuming hydrostatic equilibrium. We set the abundance of each species to be constant with height in the atmosphere. The model spectra are generated at a wavenumber spacing of 0.01~cm$^{-1}$ between 0.95-5~$\mu$m for each planet, corresponding to a spectral resolution of $\mathrm{R} = 10^6$ at 1~$\mu$m. 

The latest line lists are used to model the spectra which accurately determine the opacity for each species. We adopt the ExoMol line lists for H$_2$O \citep{polyansky2018} and NH$_3$ \citep{coles2019}, and HITEMP for CO \citep{rothman2010, li2015} and CH$_4$ \citep{hargreaves2020}. These line lists have been chosen as they are the most suitable for HRS given that they have accurately determined line positions which have have been experimentally measured and/or empirically determined from theoretical calculations. Each transition line has been spectrally broadened over a grid of pressures and temperatures using a Voigt line profile as discussed in \citet{gandhi2017}. We include H$_2$ and He pressure broadening from recent work on broadening coefficients \citep[e.g.][]{rothman2010, faure2013, barton2017_broadening}. Further details on the choice of line list and H$_2$/He broadening can be found in \citet{gandhi2020}. We also include the effect of collisionally induced absorption from H$_2$-H$_2$ and H$_2$-He interactions \citep{richard2012} and we model the effect of a cloud deck by including an additional high optical depth at pressures greater than the cloud deck pressure, P$_\mathrm{cloud}$.

\begin{figure}
	\includegraphics[width=\columnwidth,trim={0cm 0.0cm 0cm 0cm},clip]{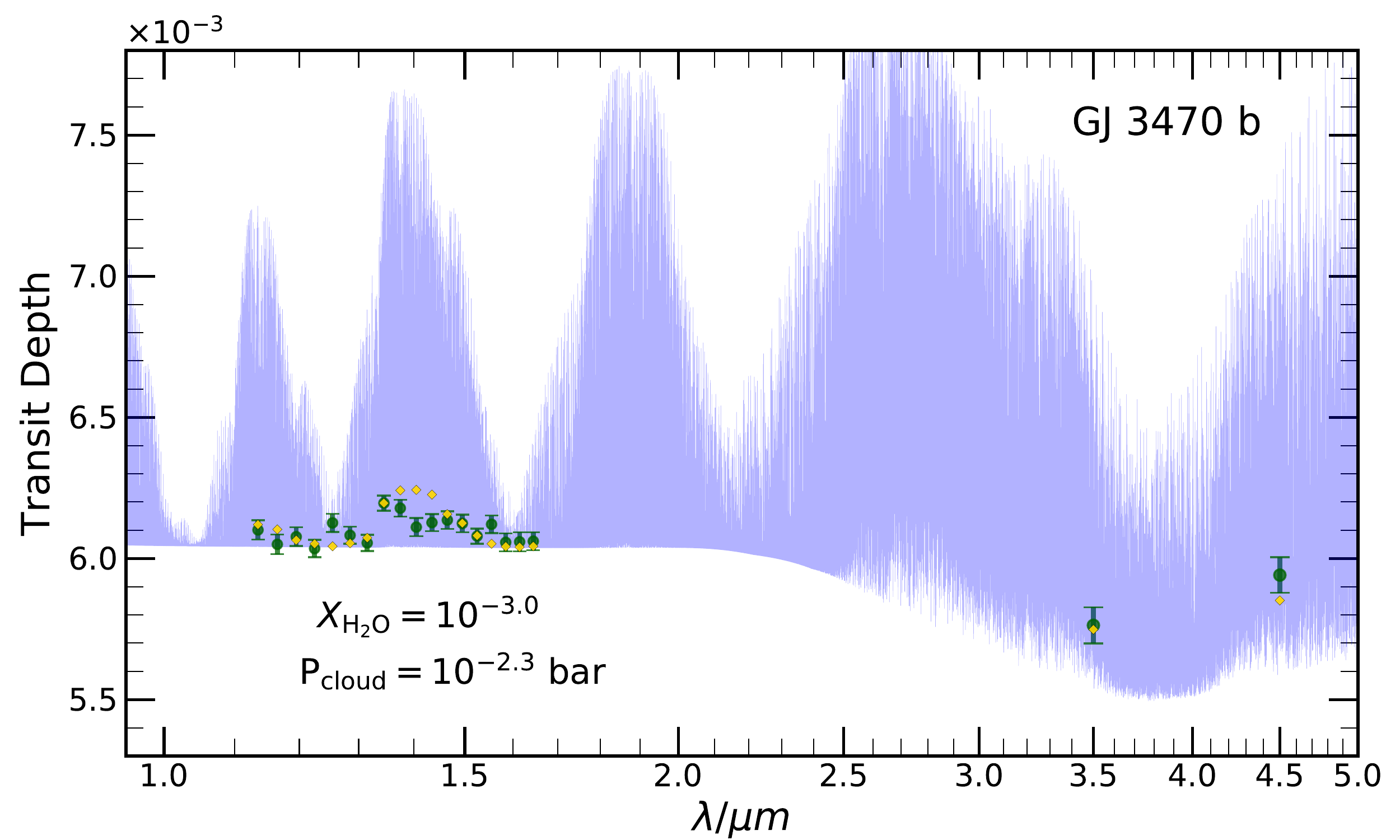}
    \caption{High resolution transmission spectrum of GJ~3470~b used to generate the simulated HRS data in Section~\ref{sec:h2o detections}. The binned WFC3 and Spitzer data points of this model are shown in yellow, and the corresponding observations \citep{benneke2019} are shown in green. We assume the atmosphere consists of an opaque cloud deck up to 2~$\mu$m whereafter the cloud opacity decreases smoothly in accordance with the low resolution constraints.}
    \label{fig:gj3470_spectrum}
\end{figure}

\subsubsection{GJ~3470~b}

The high resolution model for GJ~3470~b is derived from best-fit parameters from \citet{benneke2019} and shown in Figure~\ref{fig:gj3470_spectrum}. This model has an H$_2$O abundance of $\log_{10}(\mathrm{H_2O}) = -3.0$ and a cloud deck at $10^{-2.3}$~bar. Our cloud model prescription introduces an opacity which is constant with wavelength until a cutoff value, set to 2~$\mu$m, beyond which it smoothly decreases to 0. This is done so as to closely match the Mie scattering opacity from micron sized particulate species inferred from the low resolution HST WFC3 and Spitzer data in \citet{benneke2019}. This model is used in the simulated observations of GJ~3470~b for each of the high resolution spectrographs discussed in Section~\ref{sec:h2o detections}.

\subsubsection{GJ~436~b}\label{sec:gj436_grid}

\begin{figure*}
	\includegraphics[width=\textwidth,trim={0cm 0.0cm 0cm 0cm},clip]{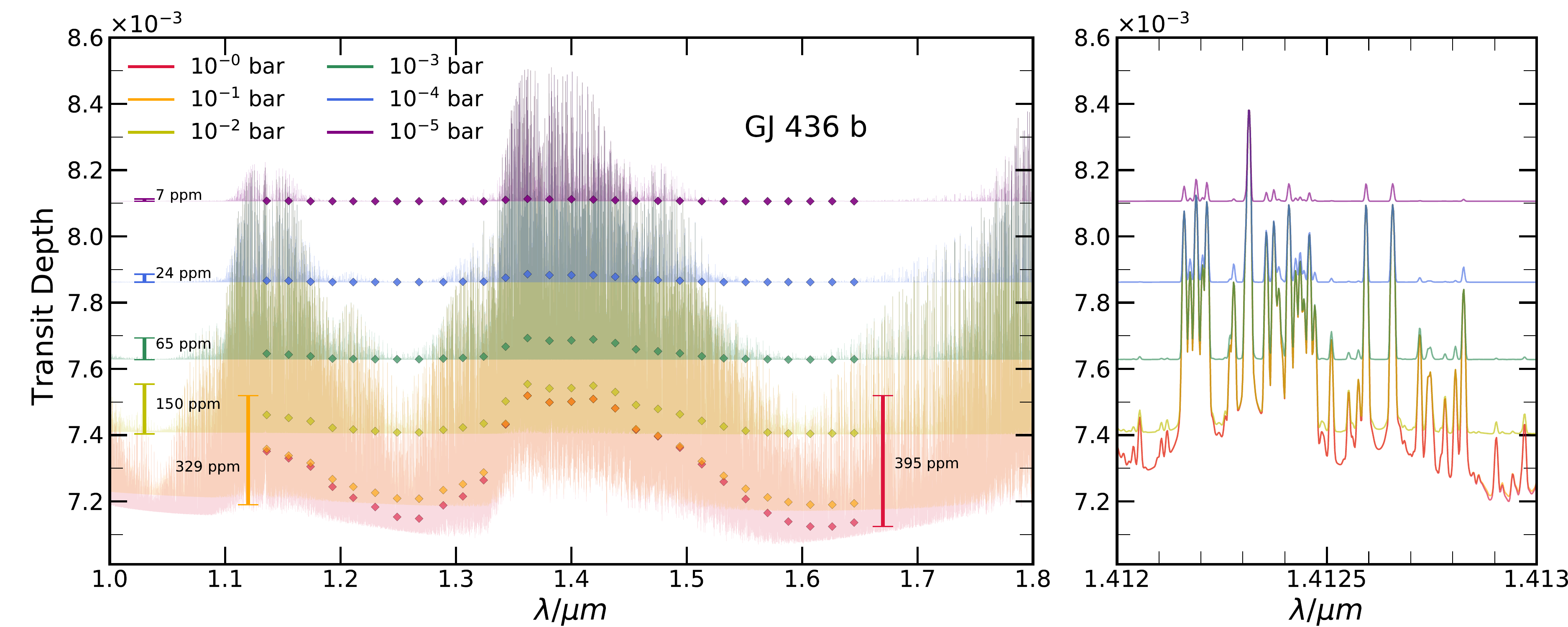}
    \caption{High resolution transmission spectrum of GJ~436~b with varying cloud deck pressures. The diamond markers in the left panel indicate the binned HST WFC3 spectrum with the corresponding height of the H$_2$O spectral feature given in parts per million. The right panel shows the high resolution spectra in a small region around 1.4125~$\mu$m. In each model the H$_2$O volume mixing ratio was fixed to $\log_{10}(\mathrm{H_2O}) = -3.25$.}
    \label{fig:gj436_wfc3_cloud_grid}
\end{figure*}

\begin{figure*}
	\centering
	\includegraphics[width=0.32\textwidth,trim={0cm 0.0cm 0cm 0cm},clip]{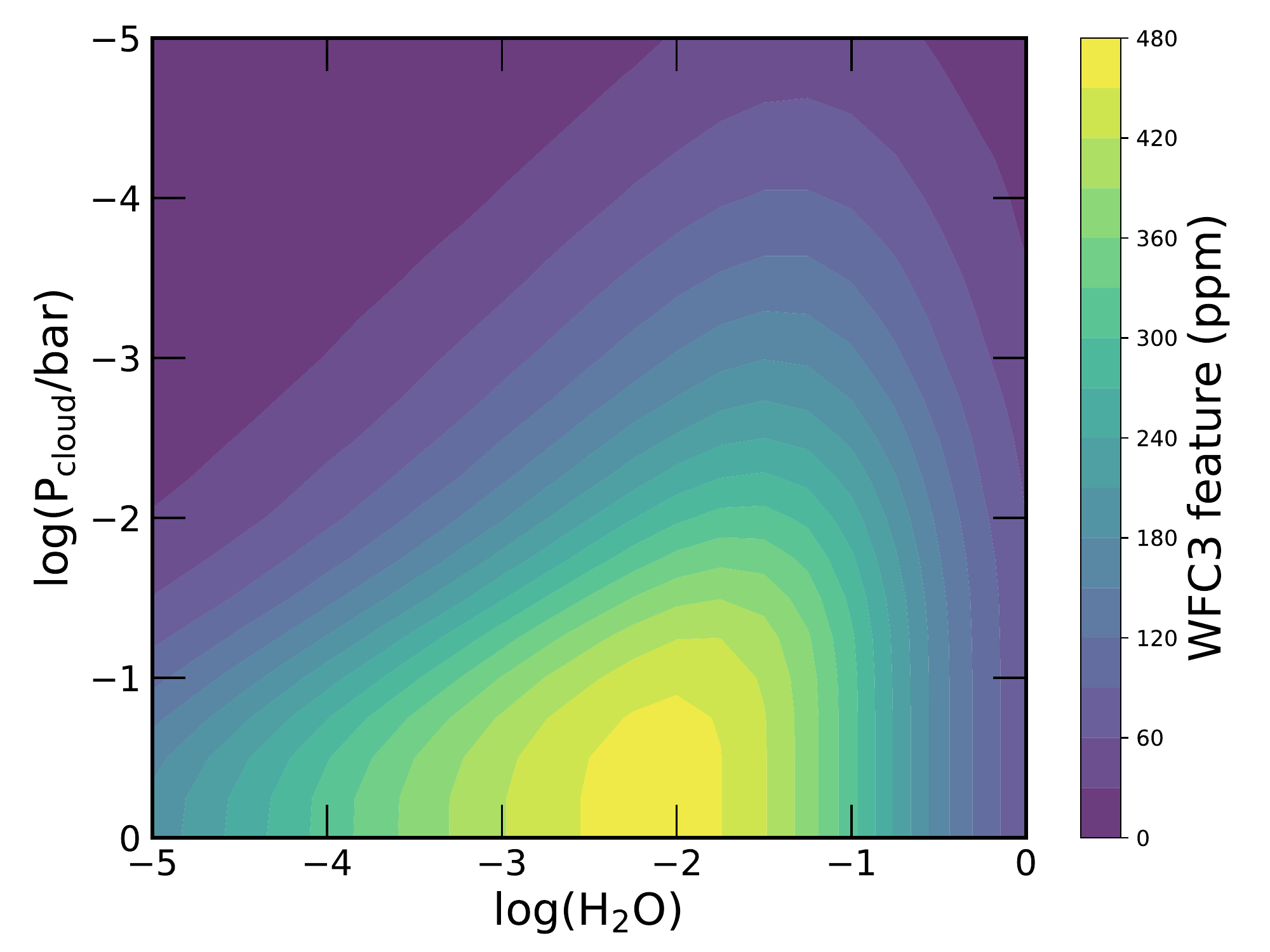}
	\includegraphics[width=0.32\textwidth,trim={0cm 0.0cm 0cm 0cm},clip]{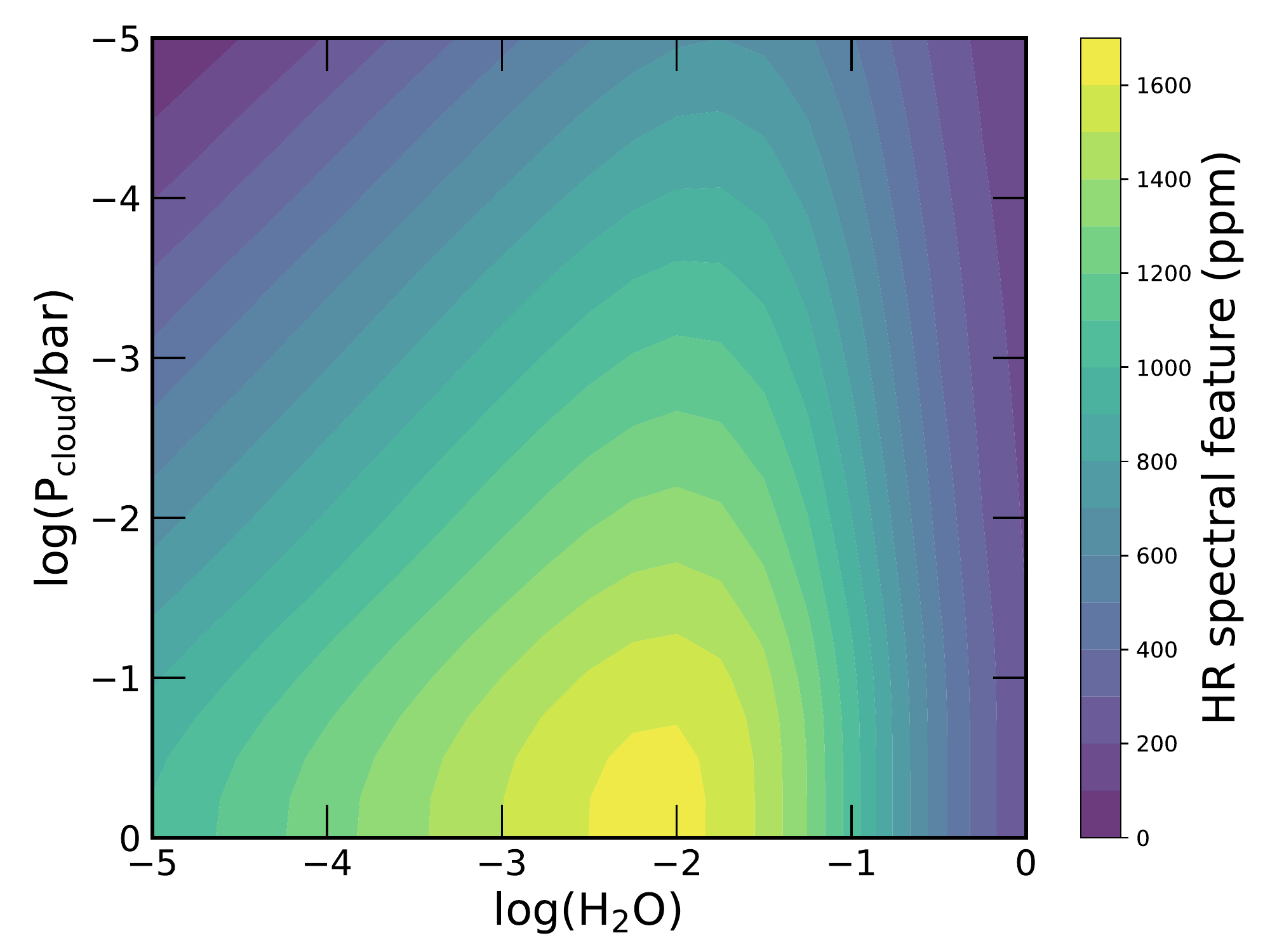}
	\includegraphics[width=0.32\textwidth,trim={0cm 0.0cm 0cm 0cm},clip]{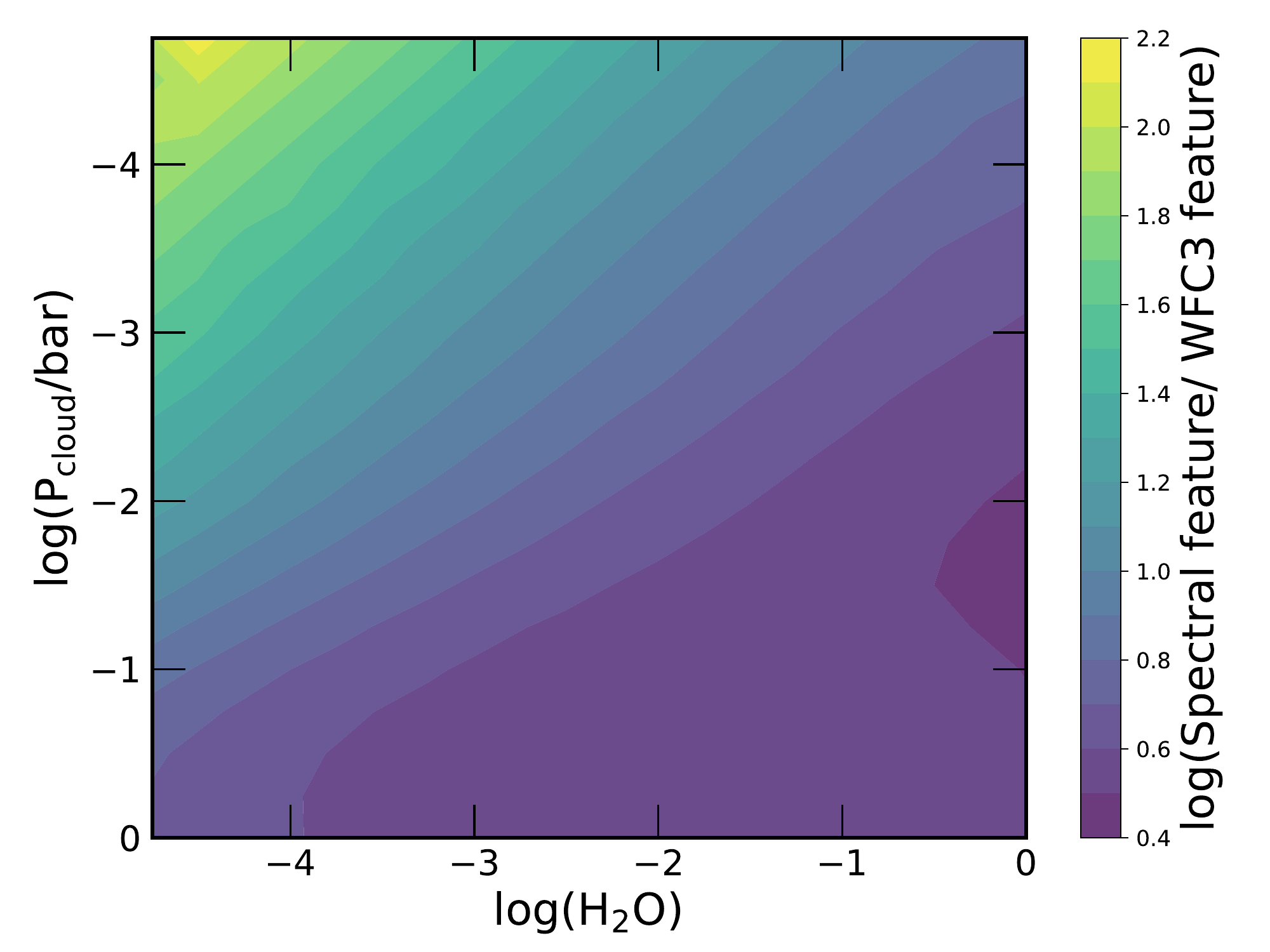}
    \caption{Spectral features in the 1.1-1.7~$\mu$m range from the grid of models for GJ~436~b discussed in Section~\ref{sec:gj436_grid}. The left panel shows the $\sim$1.4~$\mu$m H$_2$O feature in the binned HST WFC3 spectra and the middle panel shows the feature in the high resolution model spectra, with both given in parts per million. The WFC3 spectra for each model have been binned to the same resolution as the \citet{knutson2014} observations. The right panel shows the ratio of the high resolution spectral features to the WFC3 features.}
    \label{fig:features_grid}
\end{figure*}
\begin{figure*}
	\centering
	\includegraphics[width=0.49\textwidth,trim={0cm 0.0cm 0cm 0cm},clip]{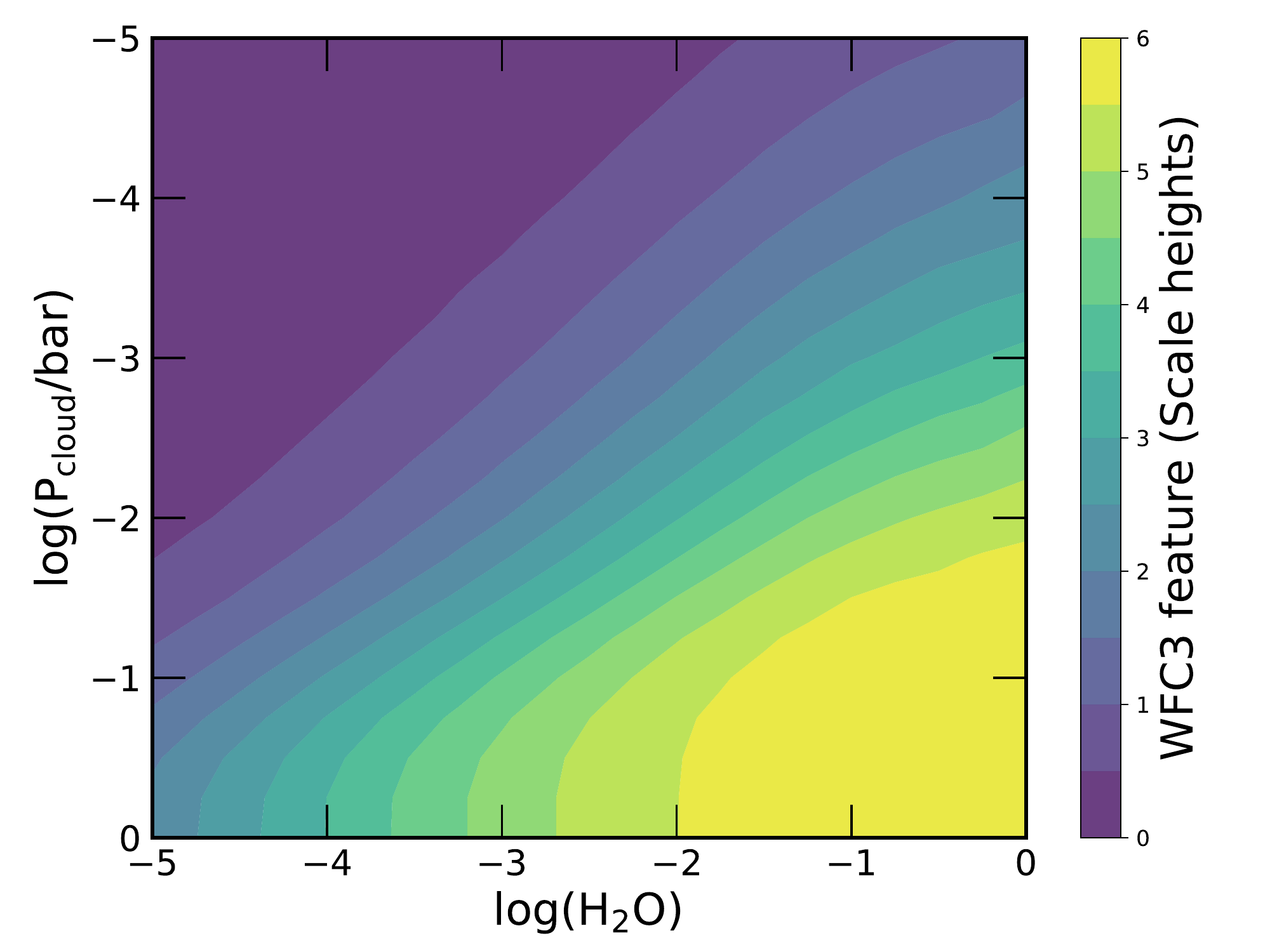}
	\includegraphics[width=0.49\textwidth,trim={0cm 0.0cm 0cm 0cm},clip]{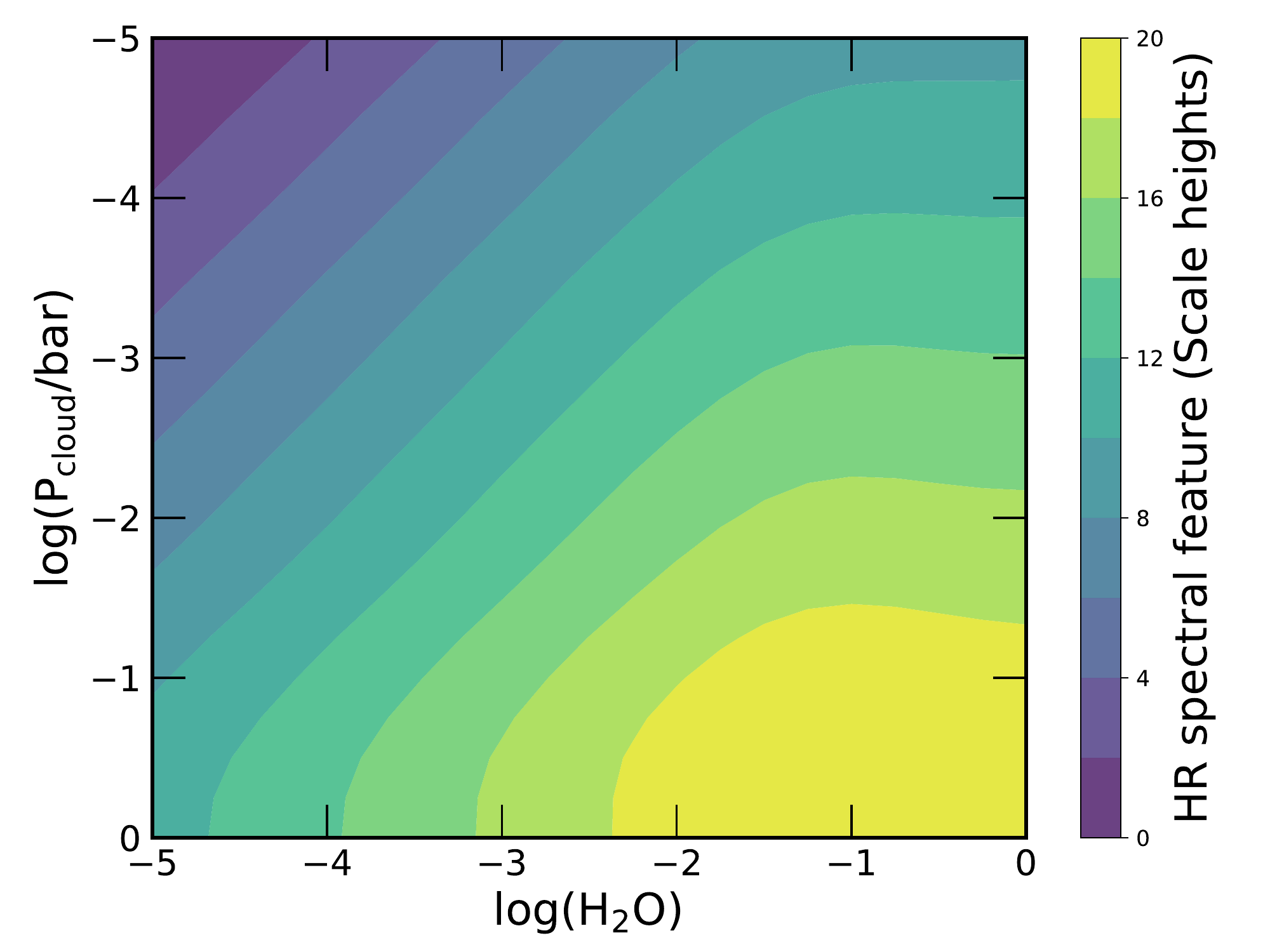}
    \caption{Spectral features in the 1.1-1.7~$\mu$m range from the grid of models for GJ~436~b discussed in Section~\ref{sec:gj436_grid}. These are given in pressure scale heights, assuming an isothermal temperature of 600~K in the atmosphere. The left panel shows the $\sim$1.4~$\mu$m H$_2$O feature in the binned HST WFC3 spectra and the right panel shows the feature in the high resolution model spectra.}
    \label{fig:features_grid_h}
\end{figure*}

We generate a grid of high resolution transmission spectra for GJ~436~b, encompassing a wide range in H$_2$O abundance and P$_\mathrm{cloud}$. This is done to explore the parameter space given that the observations of GJ~436~b revealed a nearly featureless WFC3 spectrum \citep{knutson2014} with weak constraints on the H$_2$O or cloud top pressure. This grid of spectra spans between $\log_{10}(\mathrm{H_2O}) = -5$ - 0 and $\log_{10}(\mathrm{P_{cloud}/bar}) = -5$ - 0 in steps of 0.25 dex for both parameters resulting in 441 models. We have fixed the temperature profile to be the same for each of these spectra, consistent with the equilibrium temperature. The constraints possible for HRS from this grid of models are discussed further in Section~\ref{sec:constraining_h2o_clouds_gj436b}.

Figure~\ref{fig:gj436_wfc3_cloud_grid} shows a model spectrum for GJ~436~b at solar abundance with a varying cloud deck pressure. As P$_\mathrm{cloud}$ decreases from 1~bar to 10$^{-5}$~bar, the H$_2$O spectral feature at $\sim$1.4~$\mu$m is significantly reduced in the binned HST WFC3 data. Given the observations have error bars of $\sim$50 ppm \citep{knutson2014}, this makes H$_2$O difficult to conclusively detect at P$_\mathrm{cloud} \lesssim 10^{-3}$~bar as the feature is now comparable to the error. However, the high resolution spectra (generated at 0.01~cm$^{-1}$ spacing) do have strong features even with high altitude clouds. 

We quantitatively compared the extent of the H$_2$O 1.4~$\mu$m feature over our grid of models for both the binned WFC3 and high resolution (HR) spectra as shown in Figure~\ref{fig:features_grid}. We confirm that for all H$_2$O abundances the strongest features occur for the cloud-free cases, as expected given that these have undiminished spectral lines. We also see that high altitude cloud decks show relatively stronger features in the HR spectra compared to the binned HST WFC3 data. For sub-solar H$_2$O and P$_\mathrm{cloud} \lesssim 10^{-4}$~bar, these HR features are $\sim$100$\times$ stronger than the binned HST feature. Therefore HRS may be able to detect and constrain H$_2$O even with such high clouds given that it is most sensitive to the cores of the spectral lines generated above the clouds at higher altitudes (see Section~\ref{sec:results}). This also confirms our assertions from Figure~\ref{fig:gj436_wfc3_cloud_grid} that the features remain strong above the clouds.

The HR spectra also show the strongest relative features at lower H$_2$O abundances (see Figure~\ref{fig:features_grid}). Thus HRS is comparatively more sensitive to lower abundances/trace species within the atmosphere. Additionally, for a given P$_\mathrm{cloud}$, the strongest features in the HR spectra occur for $\log_{10}(\mathrm{H_2O}) \approx -2$. However, for the cloudiest cases the binned WFC3 data the strongest features occur at higher H$_2$O abundances of $\log_{10}(\mathrm{H_2O}) \approx -1.3$. As the H$_2$O abundance is decreased below this, the spectral features in both cases begin to reduce. This is because the continuum opacity provided by the collisionally induced absorption or the cloud deck now begin to mute the H$_2$O feature in both the HR and the binned WFC3 spectrum. However, the ratio of the HR spectral features to the WFC3 features still increases.

At abundances $\log_{10}(\mathrm{H_2O}) \gtrsim -1.5$ the H$_2$O feature also decreases due to the higher mean molecular weight of an H$_2$O-rich atmosphere. The spectral features for both cases are thus reduced as the abundance of H$_2$O reaches 100\% because the atmospheric scale height, $k_bT/\mu g$, decreases by a factor of $\sim$8. Figure~\ref{fig:features_grid_h} shows the spectral features in terms of number of pressure scale heights, $k_bT/\mu g$. This removes the effect of the mean molecular weight and shows that the features remain strong at high H$_2$O abundance. The HR spectral features eventually plateau at high abundance ($\log_{10}(\mathrm{H_2O}) \gtrsim -1$) as the strongest lines begin to saturate, but the binned WFC3 features continue to increase. This is because the binned WFC3 points also have a dependence on weaker lines which continue to increase in strength as the H$_2$O abundance becomes very high. Hence at high H$_2$O abundances the ratio of the HR spectra to the binned WFC3 spectra decreases as these weaker lines become more prevalent and increase the strength of the binned WFC3 features more (see right panel Figure~\ref{fig:features_grid}).

\subsection{Model Data}\label{sec:model_data}

We generate a simulated dataset to demonstrate how HRS may be used to characterise cloudy exoplanets. We include in the simulation the essential ingredients to assess the impact of realistic sources of noise, i.e. a model for the M-dwarf star, a model telluric spectrum, and a model for instrumental efficiency, wavelength solution, and pixel scale of three near-infrared spectrographs, namely GIANO at the TNG, CARMENES at CAHA 3.5m, and SPIRou at CFHT. The characteristics of the three instruments are estimated from real data of known bright stars (HD 189733 and $\tau$ Bo\"otis) downloaded from the instrument archives, and thus provide a realistic representation of the real performances on sky.

M-dwarf model spectra are obtained from the Phoenix BT-Settl grid \citep{allard2012} and have solar metallicity. For GJ~3470, a model with surface gravity of $\log(g)=4.5$ and effective temperature of $T_\mathrm{eff} = 3600$~K is chosen. For GJ~436, a model with $\log(g)=5.0$ and $T_\mathrm{eff} = 3300$~K is chosen. These are the grid points that most closely match the stellar properties reported in the literature. Although we do not expect these M-dwarf models to accurately reproduce the position and intensity of stellar spectral lines observed at high spectral resolution, 
they will appropriately estimate the fluxes received from these stars and will reproduce the structure of their spectral bands, and thus allow us to appropriately estimate the wavelength-dependence of the signal-to-noise ratio of observations.

After converting the Phoenix spectra to SI units (W m$^{-2}$ m$^{-1}$), we compute the flux incident at the top of the Earth's atmosphere as:
\begin{equation}
    F_\mathrm{top} = F_\mathrm{model} \left( \frac{R_\mathrm{\star}}{d} \right)^2,
\end{equation}
where $R_\star$ is the stellar radius and $d$ the Sun-star distance. We adopt $R_\star = 0.510 R_\odot$, $d = 29.45$~pc for GJ~3470, and $R_\star = 0.455 R_\odot$, $d = 9.73$~pc for GJ~436. We then compute the stellar photon flux at the top of the Earth's atmosphere as
\begin{equation}
    F_\gamma = \frac{F_\mathrm{top}}{E_\gamma} = \frac{F_\mathrm{top}\lambda}{hc},
\end{equation}
where $E_\gamma = hc/\lambda$ is the average photon energy at wavelength $\lambda$, $h$ is the Planck's constant and $c$ the speed of light. 

This flux is reduced by three multiplicative terms, namely the planet transmission spectrum $T_\mathrm{P} (\lambda, t)$, the Earth's transmission spectrum $T_\oplus (\lambda)$, and the telescope-detector efficiency $\epsilon (\lambda)$. These terms all vary between 0 and 1, and are also all wavelength dependent. The planet transmission spectrum has also a temporal dependence, which is calculated as follows. Firstly, as transmission models are expressed in transit depth $\delta$, we compute the planet transmission as $T_\mathrm{P} = 1 - \delta$. The planet's spectrum is Doppler shifted according to the orbital radial velocity of the planet at each orbital phase,
\begin{equation}
    V_\mathrm{P}(t) = V_\mathrm{sys} + \frac{2\pi a}{P}\sin(i)\,\sin[2\pi\varphi(t)],
\end{equation}
which includes the systemic velocity $V_\mathrm{sys}$ and depends on the planet semi-major axis $a$, orbital period $P$, orbital inclination $i$, and orbital phase $\varphi$. In the above formula we have assumed that the orbit is circular and explicitly indicated time-dependent quantities. We have also neglected the change in barycentric velocity of the observer, since this is on the order of 100 m s$^{-1}$ during a transit. We note that in the rest of the paper we will define the planet radial velocity semi-amplitude $K_\mathrm{P}$ as
\begin{equation}
    K_\mathrm{P} = \frac{2\pi a}{P}\sin(i),
\end{equation}
that is by projecting the planet's orbital velocity along the line of sight of the observer. This formalism matches previous HRS literature.

To determine the vector of phases $\varphi(t)$ a spectral sequence centred on the mid-transit of the exoplanet is created. The exposure time is selected to be $t_\mathrm{exp} = 200$ s, which is short enough to prevent the change in radial velocity due to the orbital motion of the exoplanet to be noticeable on one single exposure, but long enough to provide enough signal to noise per spectral channel. We then include the duty cycle of each instrument, resulting in a cadence per exposure of 260 s for GIANO, 234 s for CARMENES, and 229 s for SPIRou. This results in 27-31 spectra per transit for GJ\,3470\,b (1.92-hr transit) and 14-16 spectra for GJ\,436\,b (1.0-hr transit). Lastly, the orbital phase is defined as
\begin{equation}
    \varphi(t) = \frac{t-T_\mathrm{mid}}{P},
\end{equation}
where $T_\mathrm{mid}$ is the time corresponding to the middle of the transit (i.e. the middle frame in the simulated sequence), and $P$ is the planet's orbital period, that is 3.34 and 2.64 days for GJ~3470~b and GJ~436~b, respectively.

The transmission of the Earth's atmosphere $T_\oplus(\lambda)$ is computed via the ESO Sky Model calculator for average conditions at Cerro Paranal (2.5 mm of PWV) and airmass 1.2. While these are suitable for Mauna Kea and Roque de los Muchachos (hosting SPIRou and GIANO respectively), conditions at Calar Alto (hosting CARMENES) are usually more humid, and therefore our simulations might be slightly overestimating the performances of this instrument.

We broaden $F_\gamma$, $T_\mathrm{P}$, and $T_\oplus$ by the instrument profile, i.e. by a Gaussian kernel with FWHM equal to the instrumental resolution. We then resample all these models to the wavelength solution of the spectrographs, which is also determined from real data. We compute the total amount of photons collected by one spectral channel of the simulated instruments, by accounting for the exposure time $t_\mathrm{exp}$, telescope aperture $A$, the width of the spectral channel $\Delta\lambda$, and the efficiency of the telescope+instrument system $\epsilon(\lambda)$:
\begin{equation}
    N_\gamma(\lambda,t) = F_{\gamma}(\lambda)\, T_\mathrm{P}(\lambda, t)\, T_\oplus(\lambda)\, \epsilon(\lambda)\, \Delta\lambda\,A\,t_\mathrm{exp}.
\end{equation}
Lastly, we add to each spectral channel a noise value randomly drawn from a Gaussian distribution with standard deviation $\sigma = \sqrt{N_\gamma(\lambda,t)}$. In doing so we assume that the main source of noise is Poisson noise from photon counting, which is generally a good assumptions for stars that are relatively bright in the infrared such as GJ~3470 and GJ~436. 

At this stage, if we were fully simulating a real observed spectral sequence, we would need to process the simulated data to remove the unwanted signals of the telluric and stellar spectra. As there is no consensus in the literature on the best approach to perform such a task, this stage of the analysis would render the simulations potentially dependent on the method used to process the data. 
To overcome this limitation, once the noise budget is computed and added to the data as explained above, we divide the sequence again by the modelled stellar and telluric spectra, thus effectively assuming a perfect removal of these unwanted components. We believe that while delivering a slightly optimistic result, this approach allows us to study the investigative power of HRS under different model scenarios without the risk of being biased by a specific data analysis. We defer to future work the assessment of comparison of these potential biases connected to the methodology used to process the data.

\section{Results and Discussion}\label{sec:results}

\begin{figure*}
    \centering
    \includegraphics[width=\textwidth,trim={0cm 0.5cm 0cm 0cm},clip]{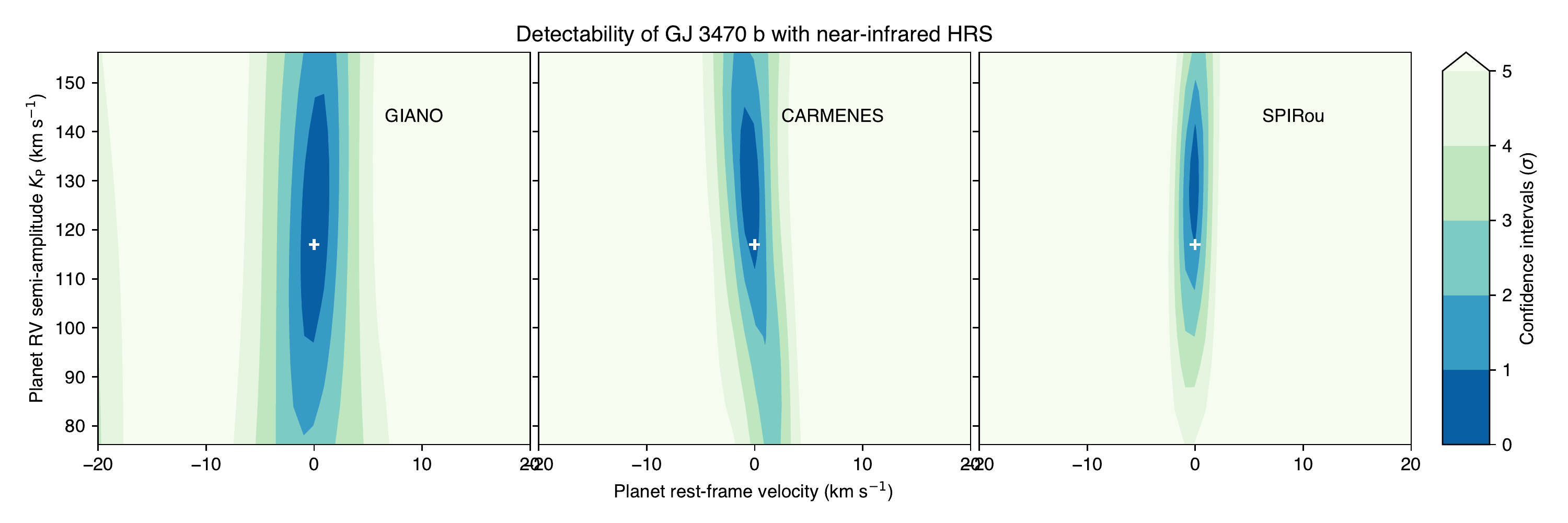}
    \caption{Detectability of the best-fitting atmospheric scenario for exoplanet GJ~3470~b as derived by \citet{benneke2019}, based on 8 hours (4 transits) of simulated observations with three near-infrared high-resolution spectrographs. Shown in the plot are confidence intervals on the rest-frame velocity and radial-velocity amplitude of the planet after the likelihood of \citet{brogi2019} has been applied to the simulated data. The white cross indicated the simulated values.}
    \label{fig:gj3470_observations}
\end{figure*}

We now discuss how HRS may be used to detect and characterise H$_2$O as well as other molecular species in the atmospheres of cloudy exoplanets. We will begin with H$_2$O on GJ~3470~b, where there is a constraint on the H$_2$O abundance from low resolution HST and Spitzer observations \citep{benneke2019}. We then explore a grid of cloudy and cloud-free spectra for GJ~436~b, a warm Neptune which has shown a cloudy atmosphere with little constraints placed on the H$_2$O abundance from HST observations \citep{knutson2014}. Finally, we explore how features for other molecular species, namely CH$_4$, NH$_3$ and CO, are affected by the presence of clouds.

\subsection{Detecting \texorpdfstring{H$_2$O}{H2O} with High Resolution Spectrographs - GJ~3470~b}\label{sec:h2o detections}

H$_2$O is one of the most important spectroscopically active molecules and one of the most well observed, present over a range of chemical compositions and temperatures \citep[e.g.][]{madhu2012, moses2013_gj436}. The model spectrum of GJ~3470~b is shown in Figure~\ref{fig:gj3470_spectrum}. Following the prescriptions of Section~\ref{sec:model_data}, we model infrared transit observations with CARMENES/CAHA, SPIRou/CFHT and GIANO/TNG. To statistically assess the properties of the planet signal, we apply the formalism of \citet{brogi2019}. We compute the log-likelihood function of each tested model by combining the data and model variances, and the cross-covariance of data and model according to their Equation~9. We assume the same total observing time as for the dataset presented in \citet{benneke2019}, which is 8.4 hours. Since HRS does not necessarily need out-of-transit data, this integration time would allow us to observe four full transits of the exoplanet. Our results are shown in Figure~\ref{fig:gj3470_observations}. We can successfully and confidently ($>5\sigma$) recover the H$_2$O signal for all instruments at the expected values of K$_\mathrm{p}$ and V$_\mathrm{sys}$, as shown by the fact that the injection values (white crosses) are within or just outside the 1-$\sigma$ confidence interval. Given the relatively strong detection achieved, we conclude that the best-fitting atmospheric scenario for GJ~3470~b would be easily detectable with current HRS with a relatively modest investment of telescope time. The size of the confidence intervals also suggests that GIANO and SPIRou are respectively the least and the most sensitive instruments based on our simulations. Note that CARMENES does not cover the $K$ band unlike the other two spectrographs but does still show a strong H$_2$O detection because of its high throughput.

\subsection{Constraining \texorpdfstring{H$_2$O}{H2O} and Cloud Decks - GJ~436~b}\label{sec:constraining_h2o_clouds_gj436b}

\begin{figure}
	\includegraphics[width=\columnwidth,trim={0cm 0.0cm 0cm 0cm},clip]{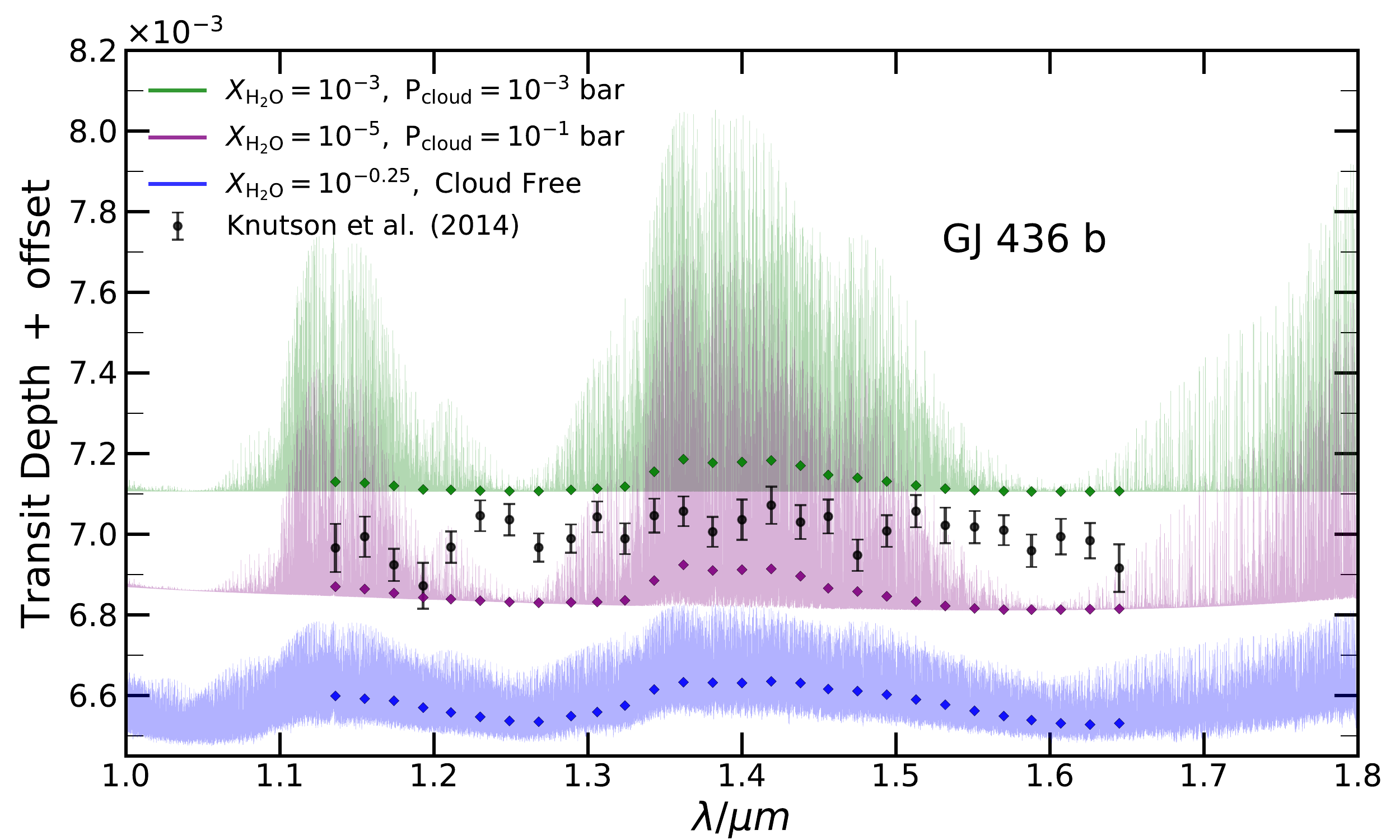}
    \caption{Model transmission spectra and binned HST WFC3 spectral points for GJ~436~b. These are shown for three cases with varying cloud pressures and H$_2$O abundances. The high resolution spectra have been binned to the \citet{knutson2014} observations, which are also shown with their associated error bars. We offset the models in the figure to highlight the spectral features.}
    \label{fig:gj436_h2o_cloud}
\end{figure}

\begin{figure*}
    \centering
    \includegraphics[width=\textwidth,trim={0cm 0.5cm 0cm 0cm},clip]{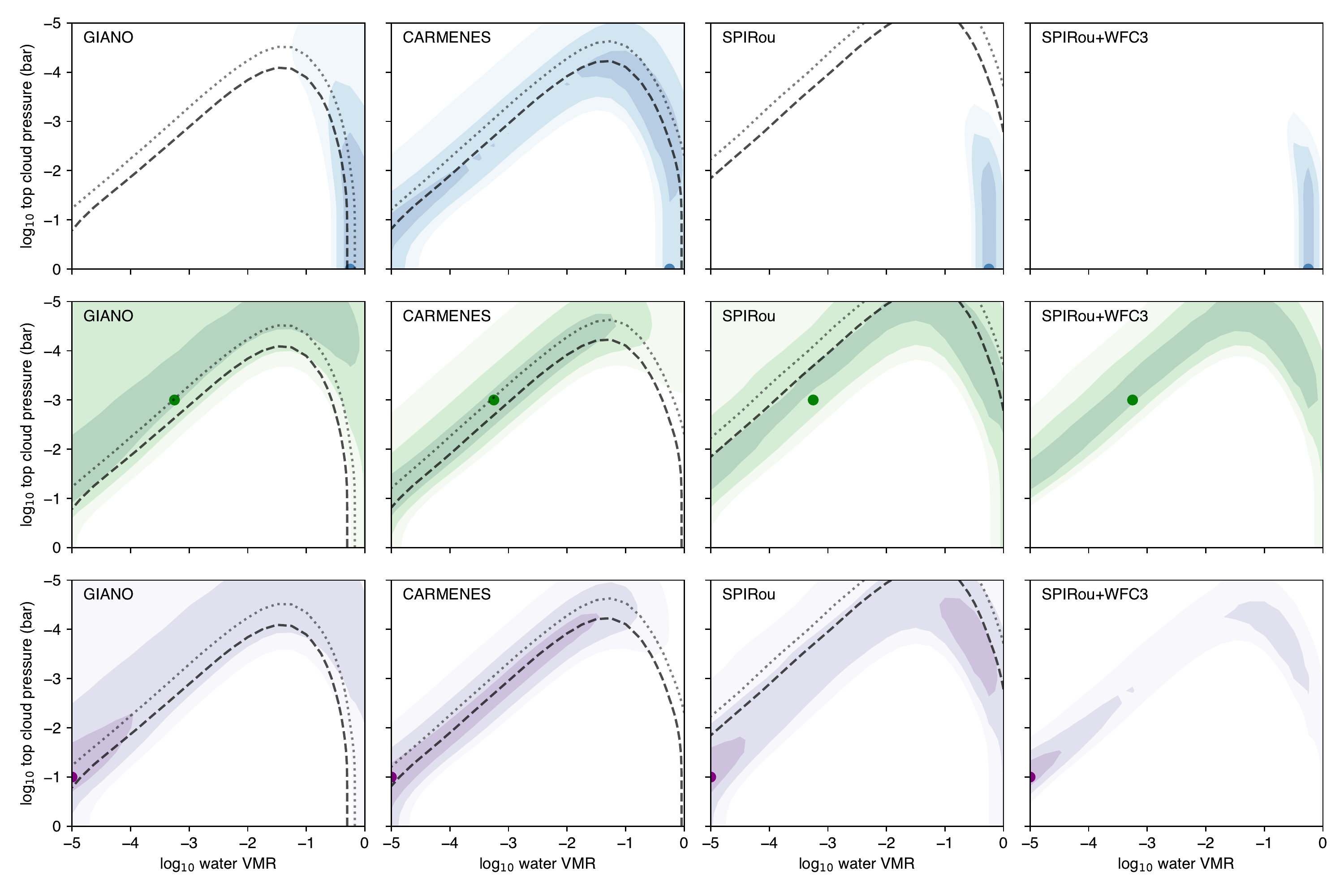}
    \caption{Simulated retrieval on GJ~436~b spectra where the abundance of water vapour and the cloud top pressure is fixed to three model scenarios (coloured dots) degenerate in low-resolution spectroscopy. These models are the same as those presented in Figure \ref{fig:gj436_h2o_cloud} and described in Section~\ref{sec:constraining_h2o_clouds_gj436b}. From left to right we present results from GIANO, CARMENES, SPIRou, and a joint simulated retrieval of SPIRou+WFC3 observations. For the three high resolution spectrographs, 3-$\sigma$ (dotted lines) and 4-$\sigma$ (dashed lines) detection limits for 10 hours of observing time are also indicated.}
    \label{fig:gj436_hrs_sim_grid}
\end{figure*}

We now discuss how HRS may be able to constrain abundances and cloud decks on GJ~436~b using the grid of model spectra discussed in Section~\ref{sec:gj436_grid}. We chose this planet as observations of the primary transit have revealed a largely featureless spectrum \citep{knutson2014}. The only constraints on H$_2$O thus far have been derived from low resolution secondary eclipse observations \citep{stevenson2010}. In Figure~\ref{fig:gj436_h2o_cloud} we show three representative high resolution transmission models with varying H$_2$O and cloud top pressures but which resulted in similar muted HST WFC3 features. We studied their detectability with CARMENES, GIANO and SPIRou, and the ability to discriminate cloudy scenarios from high-metallicity scenarios as shown in Figure~\ref{fig:gj436_wfc3_cloud_grid}.

Figure~\ref{fig:gj436_h2o_cloud} shows the three cases and the HST WFC3 observations of GJ~436~b. The binned WFC3 data for all three match the data closely and thus are degenerate and difficult to distinguish with WFC3 observations alone. On the other hand, the HRS spectrum for the H$_2$O-rich case does differ significantly to the solar and sub-solar cases, helping to partially break this degeneracy. This degeneracy can also be broken with low resolution data by the presence of non-uniform cloud cover over the terminator \citep[e.g.][]{welbanks2019} and/or observations in the optical or with Spitzer because the cloud opacity and H$_2$O opacity can vary significantly in other spectral ranges \citep[e.g.][]{benneke2019}. This is also advantageous for high resolution as spectrographs such as GIANO and SPIRou have a greater spectral coverage out to the $K$ band.

\subsubsection{Simulated Observations}

Figure~\ref{fig:gj436_hrs_sim_grid} shows the resulting confidence intervals on the H$_2$O abundance and cloud top pressure when the test models in Figure~\ref{fig:gj436_h2o_cloud} are injected in the data and retrieved against the whole grid of models. Here confidence intervals are estimated by fixing the systemic velocity and planet's orbital radial velocity to the injection values, and only exploring the H$_2$O VMR (Volume Mixing Ratio) and cloud top pressure in the retrieval. The cross-correlation-to-likelihood mapping of \citet{brogi2019} is utilised with $\Delta\log L$ values computed with respect to the best-fitting value, i.e. the model with the highest $\log L$. 

We also show the 3- and 4-$\sigma$ detection limits for 10 hours of transit observations, approximately equivalent to ten planet transits given the transit duration of 1 hour. While the number of transits observed by HST/WFC3 is only four, the total observing time is 16 HST orbits, significantly longer than these simulated observations. It is a known advantage of high-resolution cross-correlation spectroscopy that no out-of-transit baseline is required, effectively optimising the time spent on sky. In addition, these spectrographs offer higher spectral coverage in the infrared and thus are more likely to probe wavelengths with strong absorption bands from a given species. The quoted detection limits are obtained by generating a simulated dataset for each of the models in the grid, and computing its $\log L$ after cross correlating with the same model. A likelihood ratio test is then performed with the $\log L$ of a flat line, that represents the absence of signal. This is simply:
\begin{equation}
    \log L_\mathrm{flat} = -\sum_{i,j} \frac{N}{2}\log(s_{f;i,j}^2),
\end{equation}
where $s_f^2$ is the variance of the simulated spectra after their mean is subtracted out, and the sum is over each spectrum $i$ of the sequence and each order $j$ of the spectrograph.

We note that this is an exact calculation only in a simulated dataset where perfect removal of telluric and stellar spectra is assumed. It has been shown that in a real dataset imperfect removal results in data that is far from flat, hence the cross-correlation (or $\log L$) shows structures that need to be fully simulated to be properly accounted for \citep{brogi2019, buzard2020}. 

Our simulations show that for two of the three spectrographs, GIANO and CARMENES, the muted models would be only tentatively detected. This is indicated in Figure~\ref{fig:gj436_hrs_sim_grid} by the solid circles falling between 3-$\sigma$ and 4-$\sigma$ of significance. However, due to the higher overall throughput than GIANO and wider spectral range than CARMENES, SPIRou can firmly ($>4\sigma$) detect all the scenarios with the muted spectral features. 
In spite of similar performances, CARMENES shows wider confidence intervals than GIANO. This is due to the added information content of the $K$ band, which is not covered by CARMENES spectra.

We also show combined constraints which are possible using HST and SPIRou in Figure~\ref{fig:gj436_hrs_sim_grid}. These do show some improvement in the distinguishability between the cases, in particular the low and high H$_2$O abundances. However, the degeneracy remains and for the case with $\log_{10}(\mathrm{H_2O}) = -3$ and P$_\mathrm{cloud} = 10^{-3}$~bar the improvements are minimal. Hence stringent constraints on the H$_2$O abundance may still be difficult after combining low and high resolution observations unless the atmospheres have a largely cloud free atmosphere.

Despite the tentative detections of the modelled scenarios with high cloud decks, it is still possible to exclude a large fraction of the parameter space, which is qualitatively similar with the inference that is enabled by observations of flat spectra at low resolution. However, HRS allows us to go one step forward as demonstrated by the confidence intervals in Figure~\ref{fig:gj436_hrs_sim_grid}. It succeeds at discriminating atmospheres with high-metallicity from those with high cloud deck. For high-metallicity models, it is possible to correctly infer the water abundance and the low-altitude of the cloud deck. However, scenarios with lower water abundance and higher cloud deck remain highly degenerate and are poorly constrained or even biased in some instances.

\subsection{Other species}\label{sec:gj436_other_species}

As well as H$_2$O, other species also have strong spectral signatures in the infrared. At such temperatures ($\sim$500-1000~K), CH$_4$ and NH$_3$ are also expected to be present at high abundance \citep{moses2013_gj436} and are therefore important sources of opacity, along with perhaps CO at higher temperatures \citep{madhu2012}. In order to investigate whether these species have a higher chance to be detected than H$_2$O with HRS, we model additional transmission spectra of GJ~436~b with CH$_4$, NH$_3$ and CO as individual species, as well as models with all the four species mixed. We include a cloud-free case (cloud top pressure set at 1~bar), as well as progressively cloudier scenarios by placing the high-altitude cloud deck in the range $-4 < \log_{10}(P_\mathrm{cloud}) < -1$~bar, in steps of 1 dex. For these models, the abundances of the four species are fixed to $\log_{10}(\mathrm{VMR})=-3.2$ for H$_2$O, $\log_{10}(\mathrm{VMR})=-3.5$ for CH$_4$, $\log_{10}(\mathrm{VMR})=-4.2$ for NH$_3$ and $\log_{10}(\mathrm{VMR})=-6.0$ for CO. These are consistent with the solar composition chemical models by \citet{moses2013_gj436} for GJ~436~b.

\begin{figure*}
    \centering
    \includegraphics[width=5.45cm,trim={0cm 0cm 0.75cm 0cm},clip]{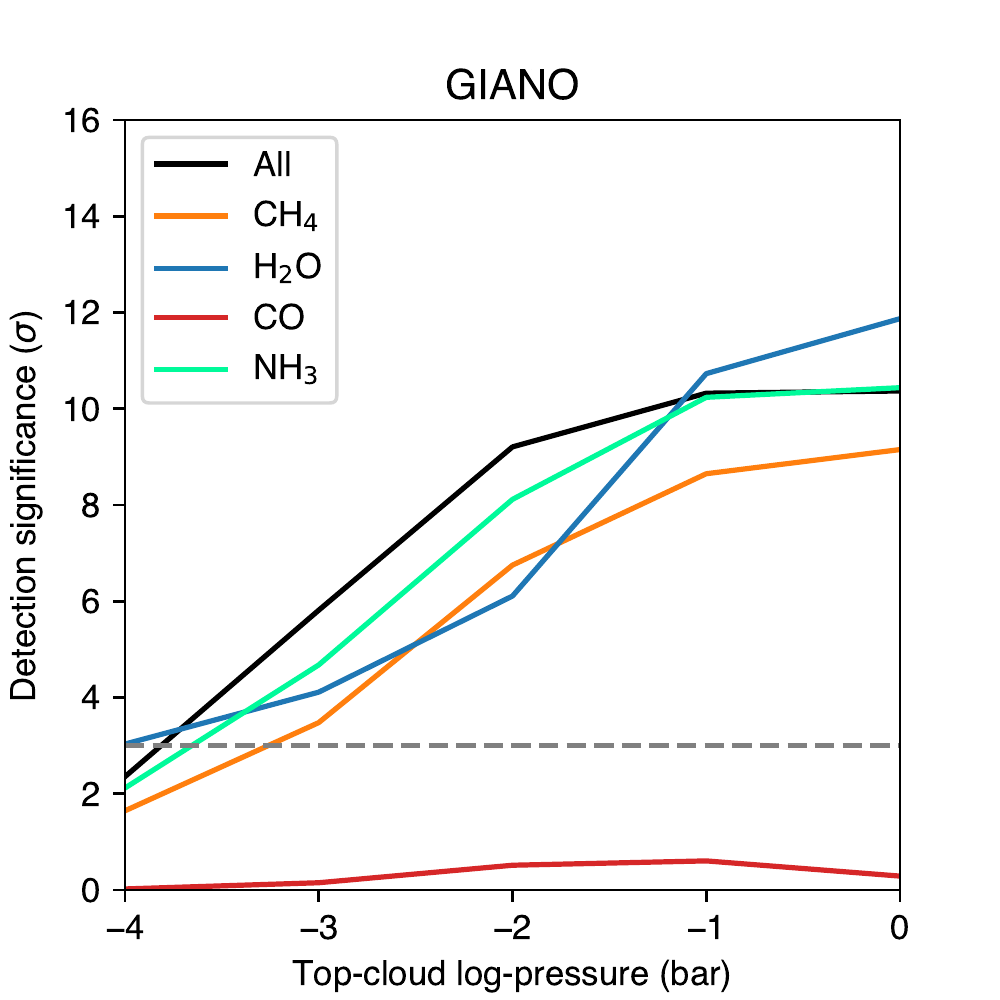}
    \includegraphics[width=5cm,trim={0.7cm 0cm 0.8cm 0cm},clip]{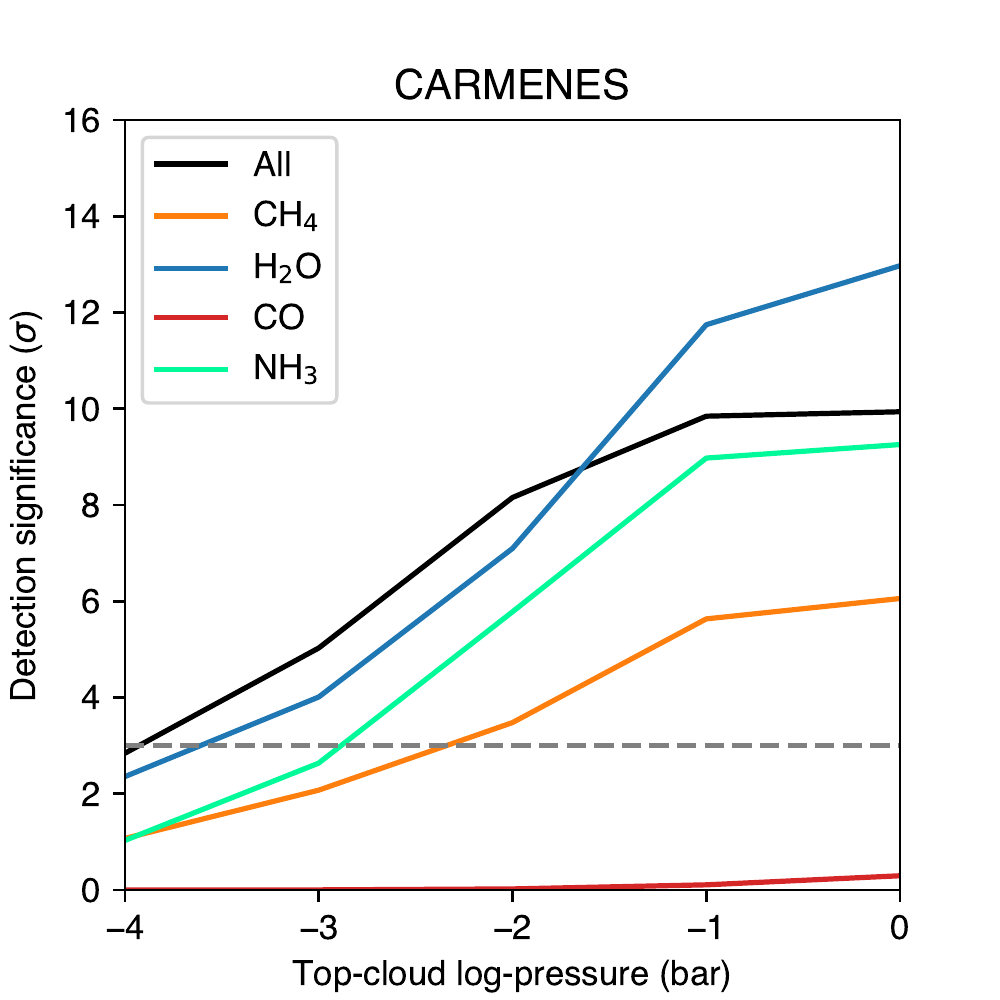}
    \includegraphics[width=5cm,trim={0.7cm 0cm 0.8cm 0cm},clip]{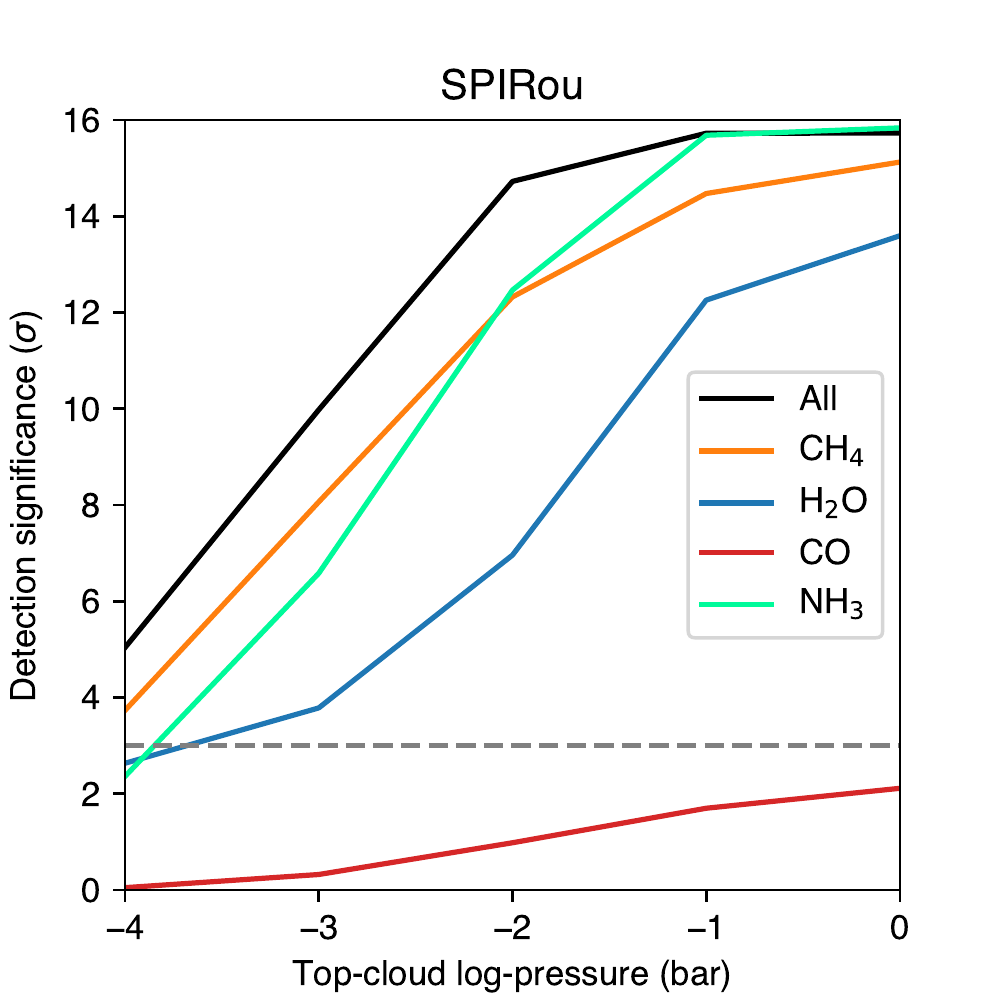}
    \caption{Simulated observations of exoplanet GJ~436~b showing the detectability of species other than water vapour (blue line) as a function of the top-cloud pressure, for 10 hours of observations with three near-infrared high-resolution spectrographs. The additional species modelled are CH$_4$ (gold), CO (red), and NH$_3$ (light green). A model containing all the species mixed together is also shown in black. The abundances are fixed to the values listed in Section~\ref{sec:gj436_other_species}. The 3$\sigma$ detection limit is indicated with a grey dashed line. It shows that all the species follow similar trends as water vapour, with differences mostly inherent to the spectral range and resolving power of the three instruments.}
    \label{fig:gj436_other_species}
\end{figure*}

We set up the simulations as in Section~\ref{sec:gj436_grid}, by assuming 10 hours of transit observations with the three infrared spectrographs. We allow each model to be retrieved not only at the injected systemic and orbital velocities, but in a wider range of $-20 < V_\mathrm{sys} < 20$ km s$^{-1}$ and $88 < K_\mathrm{P} < 166$ km s$^{-1}$, centred on the injection values of $( V_\mathrm{sys}, K_\mathrm{P})$ = (0, 128) km s$^{-1}$. This is done to simulate a real observation where we routinely verify that the significance of the detection peaks at the right values of these velocities. Indeed, in 80\% of the simulations (40 out of 50 simulations) the best-fit solution in velocity is contained within the 1-$\sigma$ confidence interval, with the remaining 20\% contained between 1 and 2$\sigma$. This is broadly in line with random fluctuations due to the noise matrix being initialised randomly for each run, and thus it points to the absence of any strong biases in the interpretation of HRS data.

The outcome of our simulations is shown in Figure~\ref{fig:gj436_other_species}. All the species studied follow a very similar trend to H$_2$O, with signals monotonically increasing for increasing cloud top pressure. The H$_2$O detection significance seems to increase more steeply for progressively clearer atmospheres, which might be a result of the fact that for very cloudy atmospheres the residual water signal overlaps with strong telluric absorption, as further discussed in Section~\ref{sec:tellurics}. Eventually the signal strength for each species begins to plateau as the cloud deck approaches 1~bar as the spectral features begin to saturate. At such high values of $P_\mathrm{cloud}$ the continuum opacity becomes dominated by the collisionally induced absorption.

\begin{figure*}
    \centering
    \includegraphics[width=17cm,trim={0cm 0.5cm 0cm 0cm},clip]{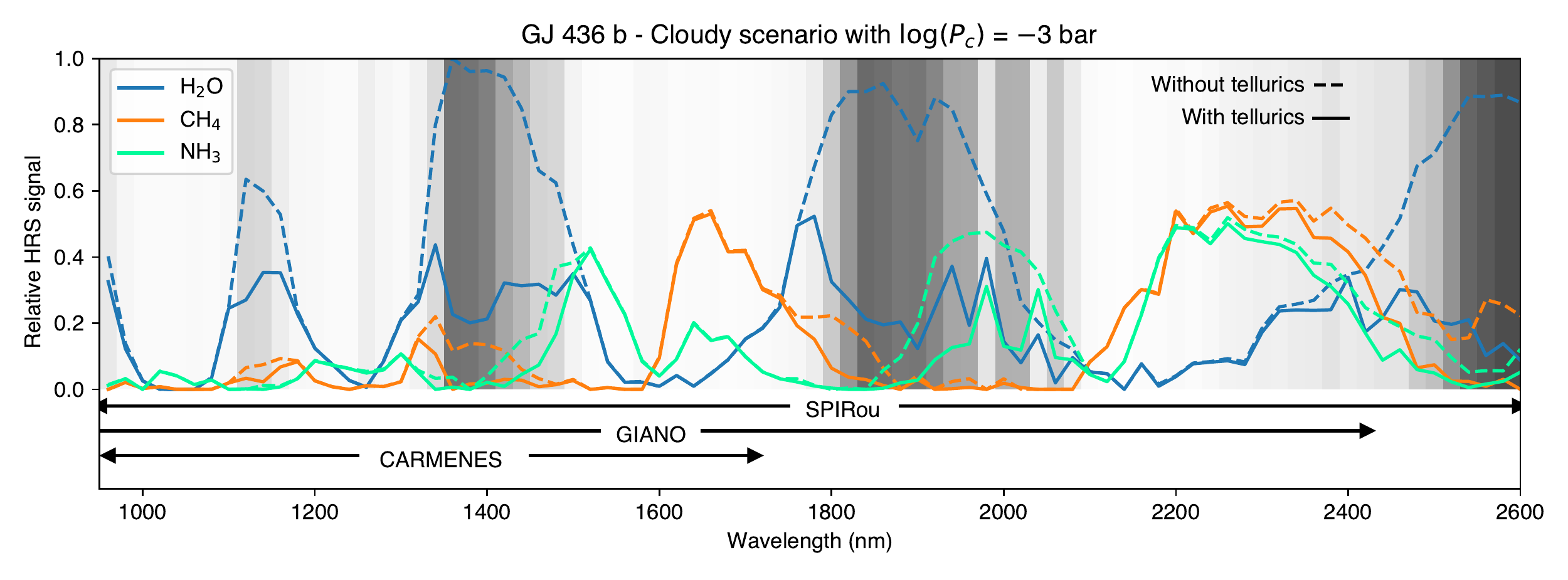}
    \caption{Relative signal content in HRS spectra as a function of wavelength, computed in 20-nm wide bins for the spectra discussed in Section~\ref{sec:gj436_other_species} and $\log_{10}(\mathrm{P_{cloud}})=-3$. Molecular species are colour-coded and their signals are plotted according to whether the effects of the Earth's atmosphere are neglected (dashed lines) or included (solid lines). The mean telluric transmission across each bin is shown by vertical bars, with brightness proportional to transmission. It shows that among the species investigated the signal of H$_2$O (in blue) is the most affected by telluric lines, not surprisingly given that the Earth's transmission spectrum is dominated by H$_2$O in the near-infrared. For reference we also show the spectral coverage of the spectrographs used for the simulations in this paper.}
    \label{fig:tellurics}
\end{figure*}

While for GIANO and SPIRou all the species except CO seem to be within the reach of these simulated observations, CARMENES struggles to detect CH$_4$ and NH$_3$ for the cloudiest scenarios (middle panel in Figure~\ref{fig:gj436_other_species}). This because both of these species have significant opacity in the $K$ band as shown in Figure~\ref{fig:tellurics}, which is not covered by CARMENES but is covered by the other two spectrographs. However, at high values of $P_\mathrm{cloud}$, H$_2$O and NH$_3$ are strongly detected by CARMENES because of its efficiency in the $H$ band where these species also have strong opacity. At the low abundance of these simulations, CO does not seem detectable by any of the spectrographs, particularly given its weak cross section everywhere except the $K$ band \citep{gandhi2020}.

It is also clear that the strength of the H$_2$O signal relative to the other species changes as a function of instrument used. For instance, while with SPIRou CH$_4$ and NH$_3$ are always detected at a higher significance than H$_2$O, with GIANO the sensitivity to the three species is approximately the same, and for CARMENES H$_2$O is always more detectable. Excluding the latter case that can again be explained by the different spectral coverage, the comparison between GIANO and SPIRou is more puzzling, and likely not reducible to just one single effect. It is important to note that, while these two spectrographs have approximately the same spectral coverage, their efficiency as a function of wavelength and their resolution differ. It is thus possible that scenarios where it is important to resolve a dense forest of lines or maximise the efficiency in certain wavelength ranges will favour SPIRou over GIANO. Resolution could also be the key to understand why with GIANO the mixed model (black line in Figure~\ref{fig:gj436_other_species}) is not detected at a higher significance than the H$_2$O model, and in fact H$_2$O alone is more easily detected for cloud-free atmospheres. When mixing species with billions of transition lines across the near infrared, blending and shielding effects between the molecules can actually reduce the line-to-line contrast. Additionally, the overall higher opacity can lift the position of the planet continuum to higher altitudes, also reducing the line-to-wing contrast. Both effects have a negative impact on cross correlation. It follows that a spectrograph with a higher resolving power should be able to reduce blending effects, thus yielding a stronger cross-correlation signal.

\subsection{Effect of Telluric \texorpdfstring{H$_2$O}{H2O} Bands}\label{sec:tellurics}

The strongest spectral features of H$_2$O are likely to be missed because of telluric absorption by H$_2$O in the Earth's atmosphere. A significant proportion of the signal is missed at $\sim1.4$~$\mu$m and $\sim1.9$~$\mu$m due to H$_2$O absorption in the Earth's atmosphere as shown in Figure~\ref{fig:tellurics}. To avoid strong telluric bands HRS observations often probe in between H$_2$O absorption bands where the opacity is weaker, which sometimes results in a band-dependent detection of water, e.g. for $\tau$ Bo\"otis b \citep[e.g.][]{brogi2012, lockwood2014}. For cloudy transmission spectra, the only significant signal that remains above the clouds is in these strong H$_2$O bands (see Figure~\ref{fig:gj436_wfc3_cloud_grid}). The detections of H$_2$O are thus weaker than expected for these cloudy cases. Whilst the current generation of spectrographs may be able to overcome this difficulty by observing a wide range of wavelengths near the strong H$_2$O features, detections with a low sigma-to-noise and/or high clouds may be more challenging in the future.

For SPIRou and some GIANO observations we can see that CH$_4$ and NH$_3$ are more detectable (see Figure~\ref{fig:gj436_other_species}), despite their lower abundance and weaker cross section than H$_2$O in the infrared \citep[see e.g.][]{gandhi2020}. This is because CH$_4$ and NH$_3$ have substantial opacity in the $H$ and $K$ bands, where telluric absorption is reduced (see Figure~\ref{fig:tellurics}). The lower abundance of CH$_4$ and NH$_3$ is able to overcome the effect of tellurics and results in an overall detectability that is higher than H$_2$O. Additionally, while the simulations in Figure~\ref{fig:tellurics} have been calculated for a fixed value of precipitable water vapour (PWV) of 2.5 mm, we recognise that different Earth observatories have broadly varying conditions. Spectrographs located at Mauna Kea, where the PWV is often lower than 2.5 mm, might actually perform better than these simulations, resulting in a better detection of H$_2$O signals. Vice-versa, sites such as Calar Alto record generally higher values than 2.5 mm, and therefore a larger fraction of the water vapour signal might be lost. Lastly, on top of the overall value of PWV, it is also important to account for its short-scale temporal variation, which is generally hard to model and de-trend with the algorithms designed for HRS so far. Residual telluric absorption generally results in time-correlated noise, and possibly spurious peaks in the final significance maps such as those shown in Figure~\ref{fig:gj3470_observations}.

As we begin to probe more Earth-like rocky planets with similar species in their atmosphere to our own planet, telluric effects will have an even more significant impact on the detectability. This reduced detectability does make the case for a space based high resolution spectrograph which could overcome telluric absorption and has the potential for strongest detections of species with prominent telluric features. As the specifics and design of space observatories can significantly differ from ground-based instrumentation, we leave the exploration of the advantages of space-based HRS to a follow up work.

\section{Conclusions}\label{sec:conclusion}

We demonstrate the feasibility of HRS to characterise the atmospheres of cloudy exoplanets with the current generation of high resolution ground based facilities. We explore how cloudy atmospheres affect the detectability and constraints on molecular species, most prominently H$_2$O. We first simulated high resolution observations of the warm sub-Neptune GJ~3470~b with the best fitting parameters from the low resolution HST and Spitzer observations \citep{benneke2019}. We show that H$_2$O is well detectable with each of the three instruments tested (GIANO, CARMENES, and SPIRou) for a modest observing time comparable to space observations. 

We then explore how HRS is able to distinguish between thick cloud and high metallicity atmospheres by modelling the atmosphere of GJ~436~b, a warm Neptune with a featureless HST WFC3 transmission spectrum \citep{knutson2014}. We focus on three representative models with varying H$_2$O and cloud but which result in similar muted features in the WFC3 range. We simulate high resolution observations with these models and cross correlate against a grid of H$_2$O abundances and cloud deck pressures. We demonstrate that a high H$_2$O abundance from a cloud-free atmosphere is distinguishable between that with clouds at solar or sub-solar H$_2$O. Hence HRS offers us the opportunity to partially break the degeneracy in WFC3 observations and characterise the atmospheres of cloudy exoplanets.

We additionally study how the detections of other trace species, namely CH$_4$, NH$_3$ and CO, vary as the cloud opacity is increased. We model the warm Neptune GJ~436~b with varying cloud to study the effect of this on the detection significance. We find that the general trend follows the water vapour signals, although trace species have a slight advantage in cloudy scenarios, due to their opacity peaking away from telluric bands.

Telluric absorption also obscures the peaks of spectral features for species such as H$_2$O which have strong opacity in the Earth's atmosphere. This may make detections with a weaker signal-to-noise more difficult in the future, particularly as we begin to characterise cooler planets more like our own. As we have shown in simulated SPIRou and GIANO observations, CH$_4$ and NH$_3$ are more detectable for cloudy atmospheres than H$_2$O, even though H$_2$O has a higher abundance and stronger cross section. This does make the case for space based high resolution spectroscopy where we would have continuous coverage over these opacity bands and thus we would have the highest potential for atmospheric characterisation of cloudy planets.

We also highlight some of the current caveats and areas of future development with atmospheric characterisation with ground based HRS. Whilst HRS is excellent for detecting trace species, standard cross correlation techniques against spectral models have historically normalised the spectra making abundance constraints difficult. However, in our work we adopt the log-likelihood metric of \citep{brogi2019}. This preserves the strength of the atmospheric spectral lines and weights the data by the noise/variance and is therefore more statistically robust for HRS applications. Retrievals with HRS observations have also recently become possible \citep{brogi2017, brogi2019, gandhi2019, Gibson2020} and will be key in reliable abundance estimates with HRS in the future.

Accurate line lists are also vital to clear detections of molecular species with HRS. This is because HRS cross correlates the cores of spectral lines and is thus sensitive to frequency shifts in line positions. \citet{brogi2019} showed the differences that line lists can produce to both the detection significances and abundance constraints. Line lists suitable for HRS have recently become available \citep[e.g.][]{rothman2010, polyansky2018, coles2019, hargreaves2020} and often use ab initio theoretical calculations with empirically determined energy levels to provide accurate line positions at such high temperatures. These line lists will maximise the cross correlation of weak spectral signatures of trace species and may make strong detections possible with HRS. The most up to date line lists for HRS can be found in \citet{gandhi2020}.

Being able to probe cloudy exoplanets may be key to detecting biosignatures on rocky worlds in the future \citep{kaltenegger2017, meadows2018}. Upcoming facilities such as ELT (Extremely Large Telescope) have shown to have the potential for such applications for nearby stars \citep{snellen2013, rodler2014, lopez-morales2019, hawker2019}. Hence understanding cloudy atmospheres at high resolution is essential for characterising cloudy rocky exoplanets.

\section*{Acknowledgements}

SG and MB acknowledge support from the UK Science and Technology Facilities Council (STFC) research grant ST/S000631/1. We thank the anonymous referee for their helpful comments on the manuscript. Posting of this manuscript on the arXiv was coordinated with C. Hood et al.

\section*{Data Availability}
The simulations and models underlying this article will be shared on reasonable request to the corresponding author.




\bibliographystyle{mnras}
\bibliography{references} 




\bsp	
\label{lastpage}
\end{document}